\documentclass{bvm} 

\makeatletter
\let\blx@rerun@biber\relax
\makeatother

\begin{document}

\newcommand{\bvmyear}{2023}

\selectlanguage{english} 

\title{Multi-Scanner Canine Cutaneous Squamous Cell Carcinoma Histopathology Dataset}

\titlerunning{Multi-Scanner Histopathology Dataset}

\author{
	Frauke \lname{Wilm} \inst{1,2}, 
	Marco \lname{Fragoso} \inst{3}, 
	Christof~A. \lname{Bertram} \inst{4}, 
	Nikolas \lname{Stathonikos} \inst{5},
        Mathias \lname{Öttl} \inst{1},
        Jingna  \lname{Qiu} \inst{2},
        Robert  \lname{Klopfleisch} \inst{3},
	Andreas \lname{Maier} \inst{1}, 
        Katharina \lname{Breininger} \inst{2, \dagger},
	Marc \lname{Aubreville} \inst{6, \dagger}
}

\authorrunning{Wilm et al.}

\institute{
\inst{1} Pattern Recognition Lab, Friedrich-Alexander-Universität Erlangen-Nünberg, Germany\\
\inst{2} Department AIBE, Friedrich-Alexander-Universität Erlangen-Nürnberg, Germany\\
\inst{3} Institute of Veterinary Pathology, Freie Universität Berlin, Germany\\
\inst{4} Institute of Pathology, University of Veterinary Medicine, Vienna, Austria\\
\inst{5} Pathology Department, University Medical Centre Utrecht, The Netherlands\\
\inst{6} Technische Hochschule Ingolstadt, Ingolstadt, Germany\\
}

\email{frauke.wilm@fau.de}

\maketitle
\renewcommand*{\thefootnote}{\fnsymbol{footnote}}
\footnotetext[2]{shared senior authors}
\renewcommand*{\thefootnote}{\arabic{footnote}}

\begin{abstract}
In histopathology, scanner-induced domain shifts are known to impede the performance of trained neural networks when tested on unseen data. Multi-domain pre-training or dedicated domain-generalization techniques can help to develop domain-agnostic algorithms. For this, multi-scanner datasets with a high variety of slide scanning systems are highly desirable. We present a publicly available multi-scanner dataset of canine cutaneous squamous cell carcinoma histopathology images, composed of 44 samples digitized with five slide scanners. This dataset provides local correspondences between images and thereby isolates the scanner-induced domain shift from other inherent, e.g. morphology-induced domain shifts. To highlight scanner differences, we present a detailed evaluation of color distributions, sharpness, and contrast of the individual scanner subsets. Additionally, to quantify the inherent scanner-induced domain shift, we train a tumor segmentation network on each scanner subset and evaluate the performance both in- and cross-domain. We achieve a class-averaged in-domain intersection over union coefficient of up to 0.86 and observe a cross-domain performance decrease of up to 0.38, which confirms the inherent domain shift of the presented dataset and its negative impact on the performance of deep neural networks.        
\end{abstract}

\section{Introduction}
\label{3337-sec-introduction}
Digitizing histological specimens with dedicated slide scanning systems has facilitated machine learning-based image analysis for histopathology. These algorithms have since assisted pathologists in a variety of routine tasks, e.g. mitotic figure detection \cite{3337-01}, for which they have been able to outperform trained experts in controlled settings \cite{3337-01, 3337-02}. Still, their performance is highly dependent on the quality and availability of training data \cite{3337-03} and can deteriorate considerably on a test set where the image characteristics differ from the training data \cite{3337-04}. Such differences commonly referred to as \elqq domain shift\erqq{} can originate not only from different staining and tissue preparation protocols of different pathology laboratories but also from the digitization of histological specimens with different scanning systems. Especially from a clinical perspective, domain-agnostic models are important for generating accurate and reliable predictions.   

Previous work has shown that domain generalization techniques, e.g. domain-adversarial training, can help to develop domain-agnostic models \cite{3337-05}. For this, a training dataset composed of a wide range of different domains is highly desirable. So far, the most extensive publicly available multi-scanner histopathology dataset is the training set of the MICCAI MItosis DOmain Generalization (MIDOG) 2021 challenge \cite{3337-02}. The dataset consists of $2\ts \text{mm}^2$-sized cropped regions of 200 breast cancer cases digitized with four scanners. However, the cases were divided between the scanners, and performance differences can therefore not solely be attributed to the slide scanner but also to the case selection. The Mitos \& Atypia dataset \cite{3337-06} is the only public multi-scanner histopathology dataset with local image correspondences, i.e. the same case was digitized with multiple slide scanners, however, with 16 cases and two scanners, its extent is limited and it does not leave room for experiments with hold-out test scanners.        

In this work, we present a canine cutaneous histopathology dataset, where each of the 44 samples was digitized with five different slide scanning systems. This multi-scanner dataset provides local image correspondences, useful for domain generalization experiments. Accompanied by an annotation database of 1,243 polygon annotations for seven histologic classes (tumor, epidermis, dermis, subcutis, bone, cartilage, and a combined class of inflammation and necrosis), this is the first publicly available multi-scanner segmentation dataset. For each scanner subset, we provide a detailed evaluation of color distributions, sharpness, and contrast. To quantify the extent of the scanner-induced domain shift, we performed a technical validation of the dataset by training a baseline tumor segmentation algorithm on each single scanner domain and then testing the algorithm across all scanners. For some scanners, we observed a considerable performance decrease, which highlights the domain shift inherent in the dataset. The whole slide images (WSIs) and annotation databases are publicly available on Zenodo (\url{https://doi.org/10.5281/zenodo.7418555}), and code for implementing the baseline architectures can be obtained from our GitHub repository (\url{https://github.com/DeepPathology/MultiScanner_SCC}).       

\section{Materials and methods}
\label{3337-sec-mat_met}
\begin{figure}[t]
    \centering
    \setlength{\figwidth}{0.19\textwidth}
\caption{Exemplary region of interest of the multi-scanner dataset.}
\label{3337-fig-01}
    \begin{subfigure}{\figwidth} 
        \includegraphics[width=\textwidth]{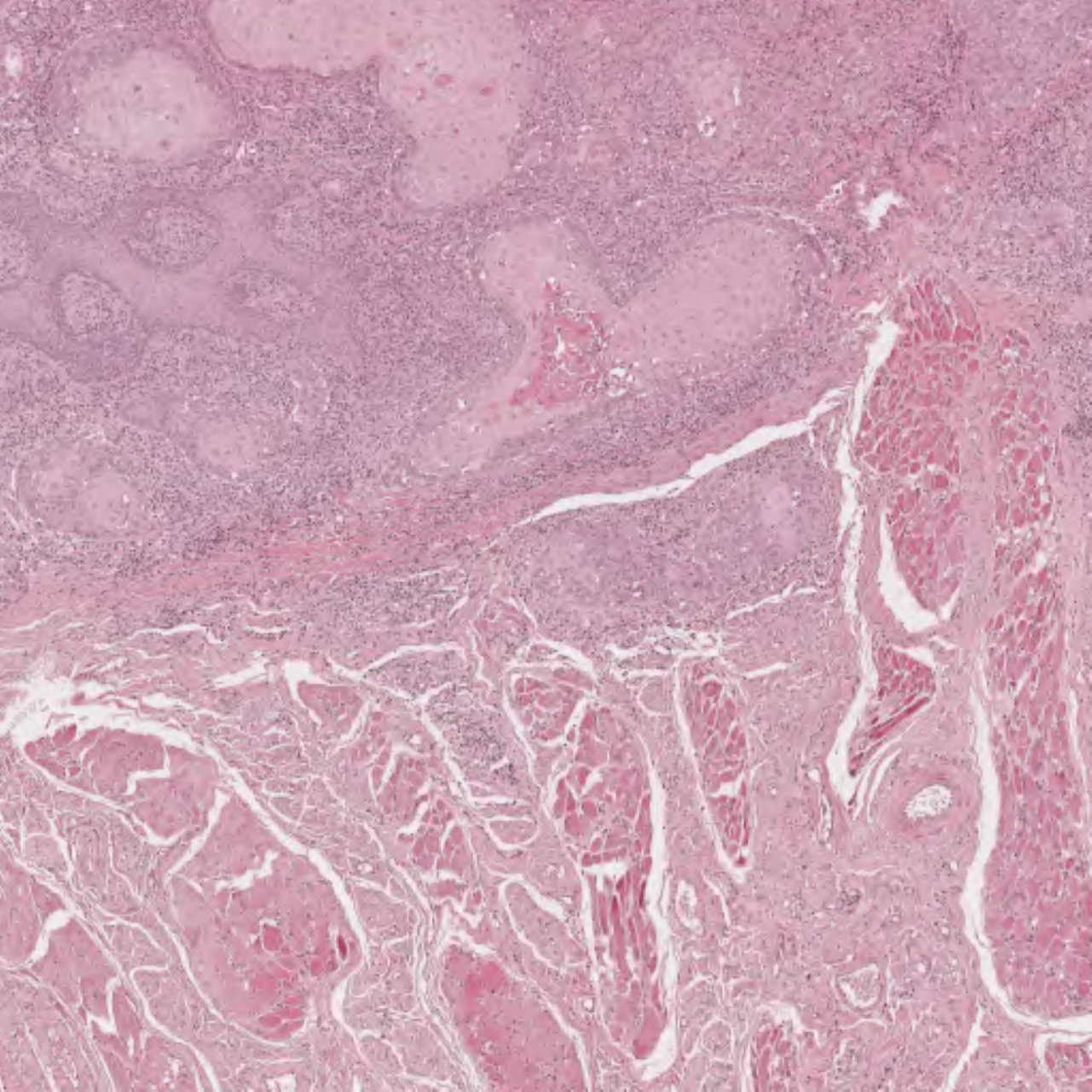} 
	\caption{CS2}
	\label{3337-fig-01-a}
    \end{subfigure}
    \hfill	
    \begin{subfigure}{\figwidth}
        \includegraphics[width=\textwidth]{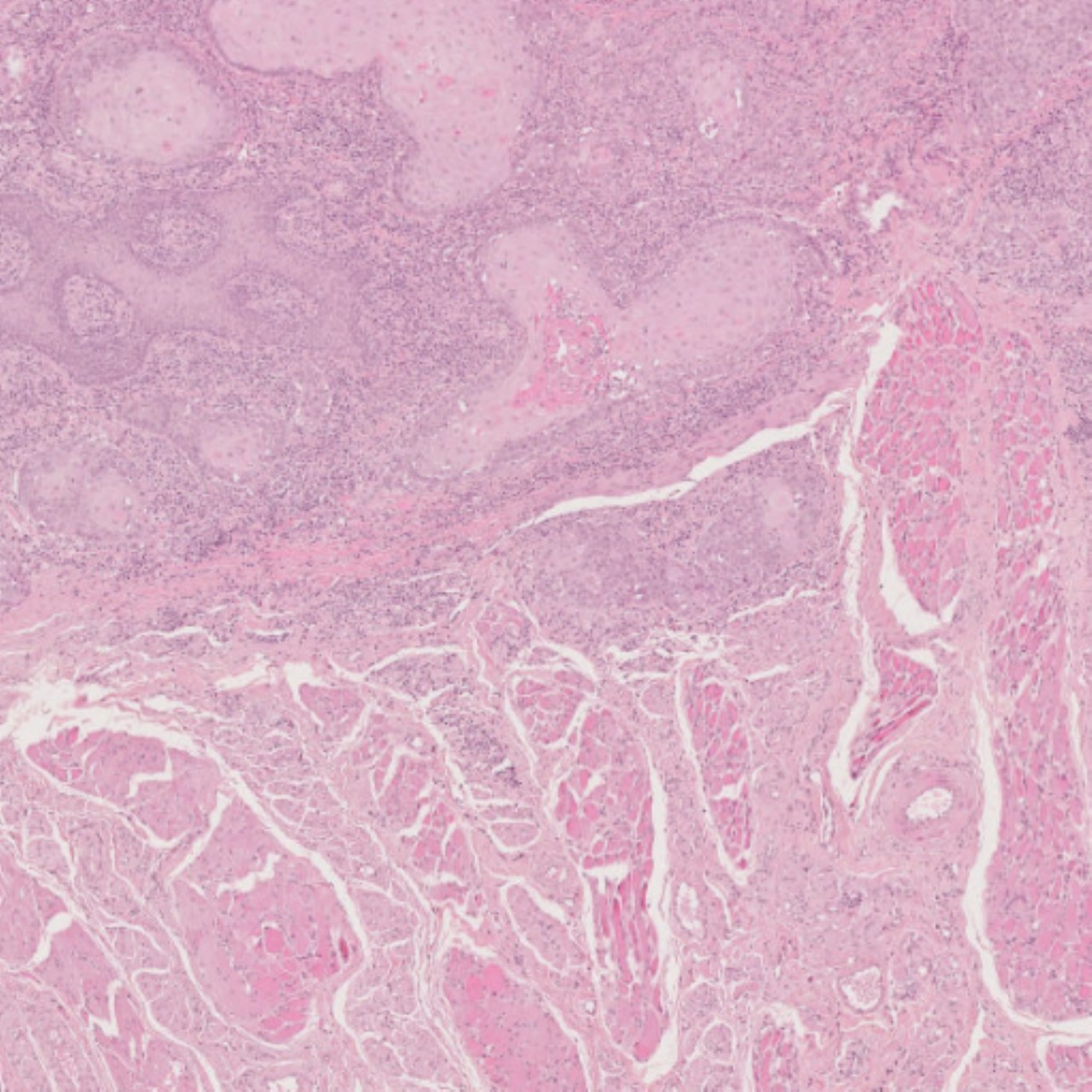}
        \caption{NZ210}
        \label{3337-fig-01-b}
    \end{subfigure}
    \hfill
    \begin{subfigure}{\figwidth}
        \includegraphics[width=\textwidth]{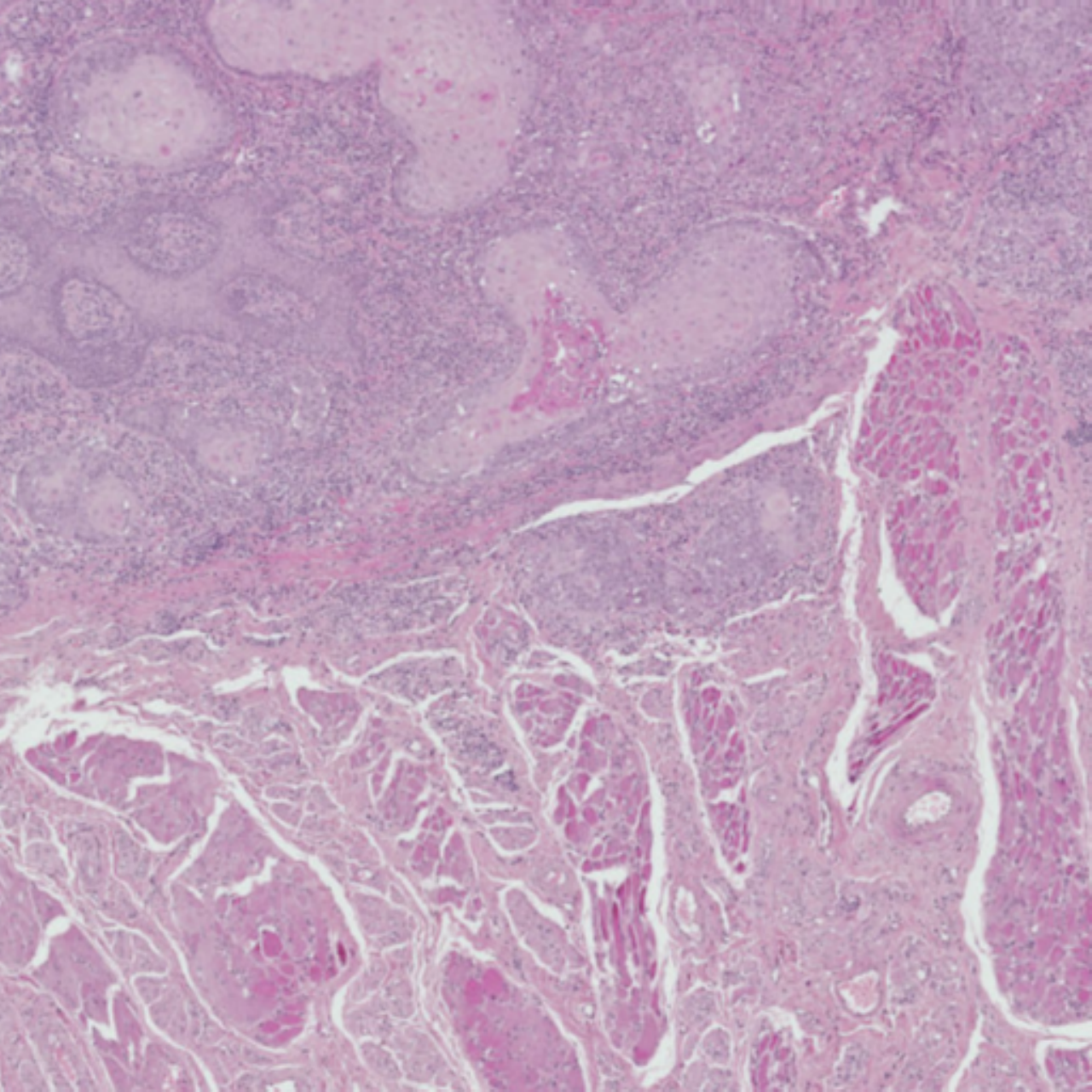}
        \caption{NZ2.0}
        \label{3337-fig-01-c}
    \end{subfigure} 
    \hfill
    \begin{subfigure}{\figwidth}
        \includegraphics[width=\textwidth]{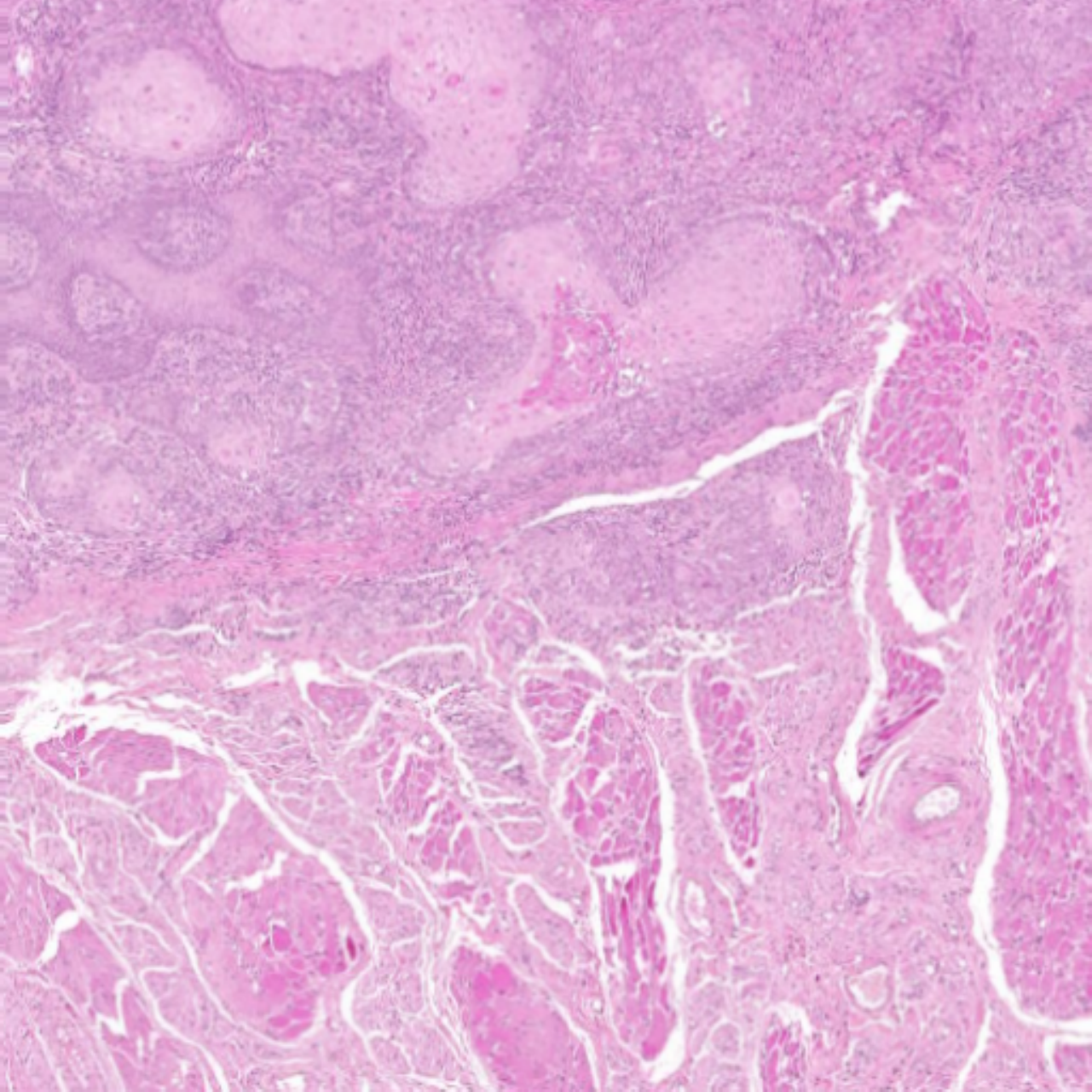}
        \caption{P1000}
        \label{3337-fig-01-d}
    \end{subfigure} 
    \hfill
    \begin{subfigure}{\figwidth}
        \includegraphics[width=\textwidth]{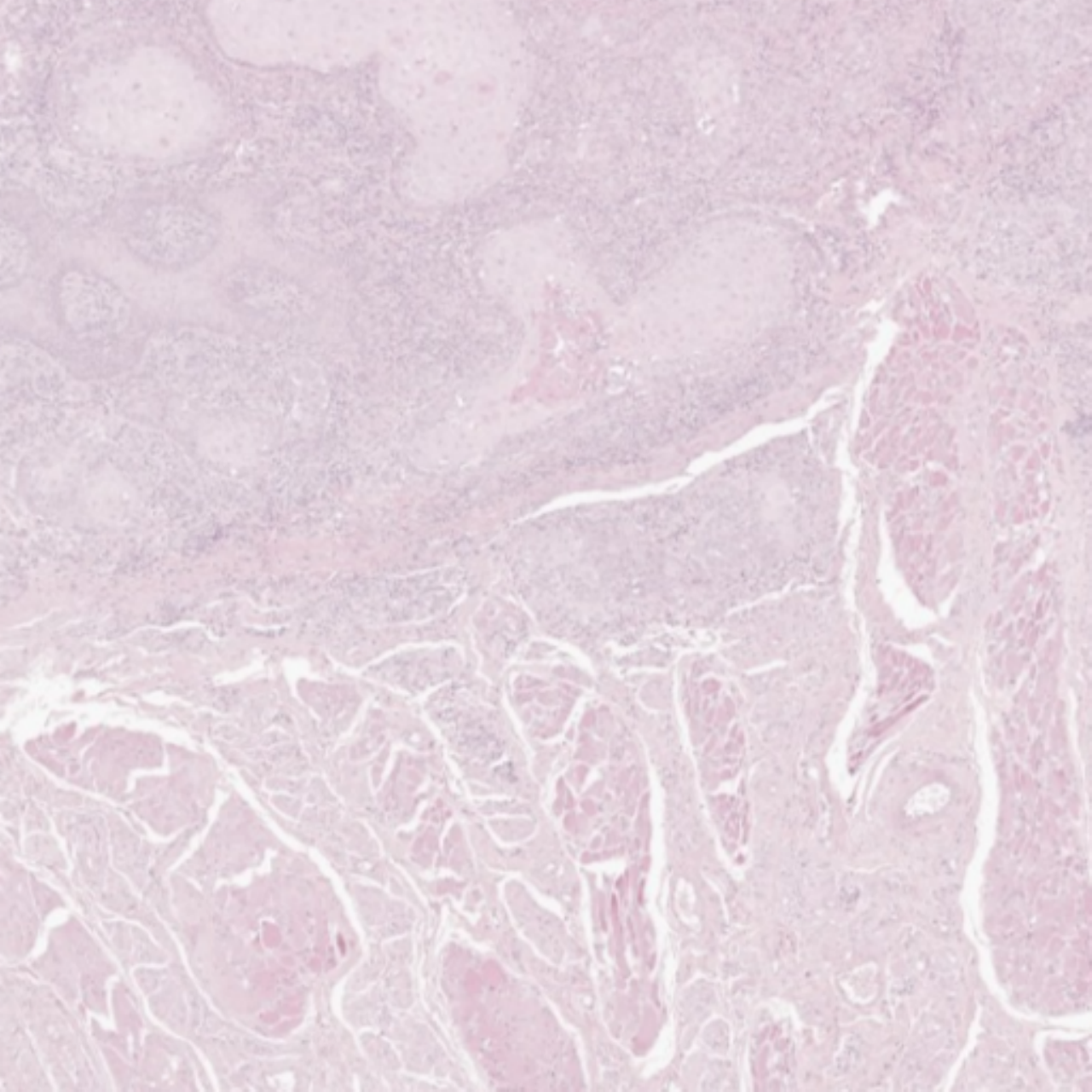}
        \caption{GT450}
        \label{3337-fig-01-e}
    \end{subfigure} 
\end{figure}

The dataset presented in this work extends the publicly available CATCH dataset \cite{3337-07}, a collection of 350 WSIs of seven of the most common canine cutaneous tumor subtypes (50 WSIs per subtype). For the CATCH dataset, the specimens were digitized with the Aperio ScanScope CS2 (Leica, Germany) at a resolution of $0.25\ts \text{µm/pixel}$ using a $40\ts \times$ objective lens. Use of these samples was approved by the local governmental authorities (State Office of Health and Social Affairs of Berlin, approval ID: StN 011/20). For the multi-scanner dataset, we randomly selected one subtype (squamous cell carcinoma) and digitized the samples with four additional slide scanners (Fig.~\ref{3337-fig-01}): 
\begin{itemize}
    \item NanoZoomer S210 (Hamamatsu, Japan), $0.22\ts \text{µm/pixel}$
    \item NanoZoomer 2.0-HT (Hamamatsu, Japan), $0.23\ts \text{µm/pixel}$
    \item Pannoramic 1000 (3DHISTECH, Hungary), $0.25\ts \text{µm/pixel}$
    \item Aperio GT 450 (Leica, Germany), $0.26\ts \text{µm/pixel}$ 
\end{itemize}%

Due to severe scanning artifacts in at least one of the scans, six specimens were excluded from the dataset, resulting in a total of 220 WSIs (44 samples digitized with five scanners each). The CATCH annotation database provides annotations for the individual tumor subtypes and six additional skin tissue classes (epidermis, dermis, subcutis, bone, cartilage, and a combined class of inflammation and necrosis). We transferred all annotations to the other scanners using the WSI registration algorithm by Marzahl \etal~\cite{3337-08} and visually validated them by overlaying the transformed polygon annotations onto the scans. We provide public access to the WSIs on Zenodo (\url{https://doi.org/10.5281/zenodo.7418555}), licensed under a Creative Commons Attribution 4.0 International License. However, due to storage restrictions, we have converted them to lower-resolution pyramidal TIFFs ($4\ts \text{µm/pixel}$), which has shown to be adequate for training segmentation tasks on the CATCH dataset \cite{3337-09}.

\subsection{Dataset validation}
For each scanner subset, we evaluated the average RGB color distribution, sharpness, and contrast. For sharpness estimation, we used the cumulative probability of blur detection (CPBD) metric \cite{3337-10}, which is a perceptual-based image sharpness metric. It is computed via edge detection, followed by a blur estimation at the detected edges. The CPBD metric then corresponds to the cumulative probability of blur detection, i.e. the percentage of image edges that fall below a threshold of a perceptually noticeable blur. For implementation details, we refer to \cite{3337-10}. For the analysis of RGB distributions and contrast, we used Otsu's adaptive thresholding to separate foreground tissue from white background. For each image, we calculated the average intensities of the color channels $I_R, I_G,\text{ and }I_B$ in the detected tissue regions. Afterward, we converted the regions to grayscale and computed the Michelson contrast \cite{3337-11} $C_M$ as a measure of global contrast.

\subsection{Technical validation}
For technical validation of the dataset, we trained a segmentation model on each scanner domain and tested the algorithm across all scanners. For model development, we performed a slide-level split into training (N=30), validation (N=5), and test (N=9) cases. We trained a UNet with a ResNet18 encoder pre-trained on ImageNet for the segmentation into tumor, non-tumor, and background. For this, we combined all skin tissue classes into one non-tumor class and used the automatically detected background areas to train the background class. We trained the networks on image patches sized 512\ts$\times$\ts512 pixels, extracted at a resolution of $4\ts \text{µm/pixel}$. During each epoch, we sampled 50 patches per WSI within the annotated polygons. Due to a high class imbalance, we randomly sampled the polygons with a class-weighting of 10\ts\% background and 45\ts\% each of tumor and non-tumor regions. For each scanner, we applied z-score normalization with the training set statistics (mean and standard deviation) and performed data augmentation using random flipping, affine transformations, and random lightning and contrast change. We used the Adam optimizer and trained the networks with a combination of cross-entropy and Dice loss. We trained the models with a batch size of 8 and a cyclic learning rate of $10^{-4}$ for 100 epochs, after which we observed model convergence. Model selection was guided by the highest intersection over union (mIoU) on the validation set.  

\section{Results}
\label{3337-sec-results}
\begin{figure}[b]
    \centering
    \setlength{\figwidth}{0.19\textwidth}
    \begin{subfigure}{\figwidth} 
        \includegraphics[width=\textwidth]{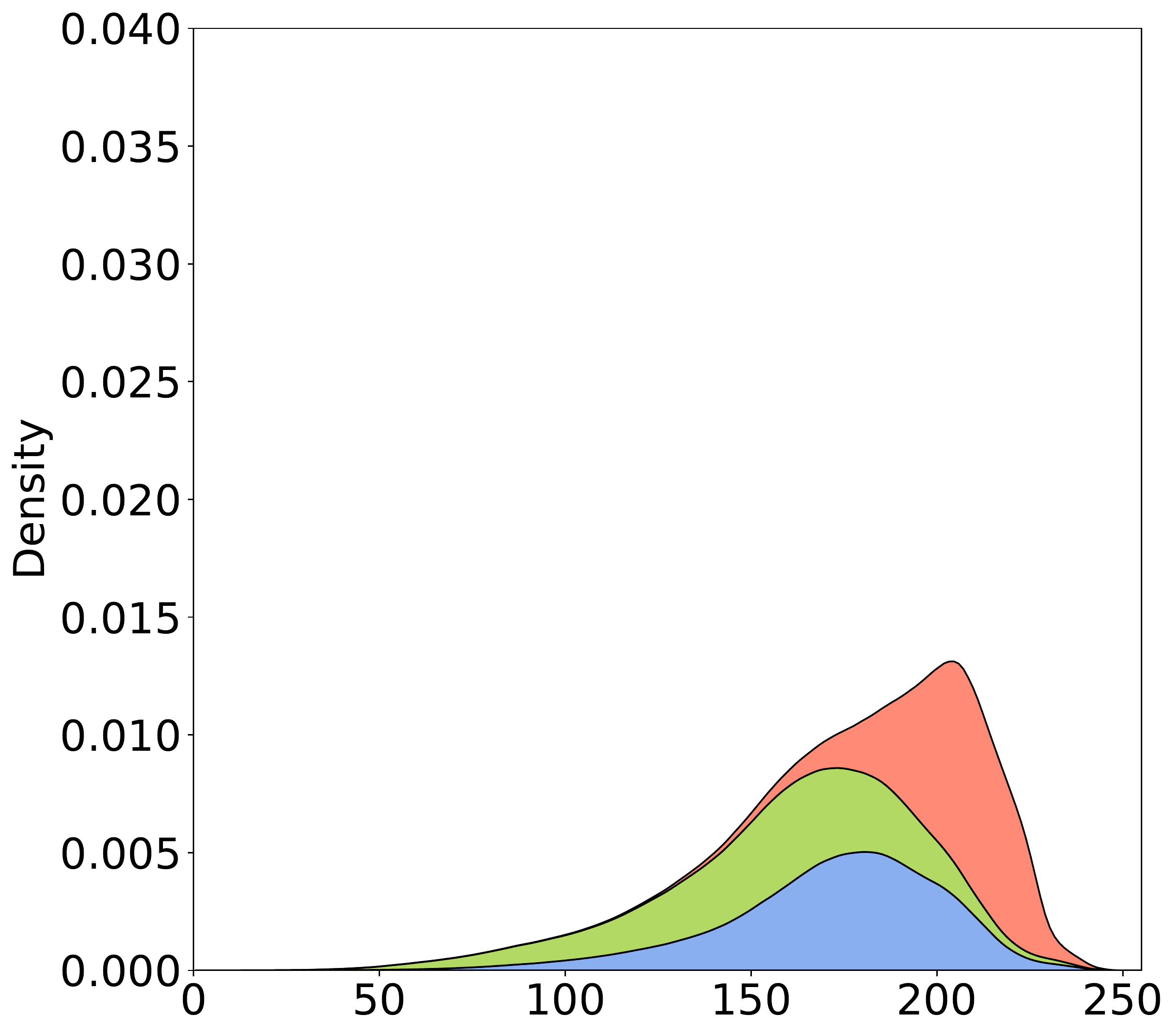} 
	\caption{CS2}
	\label{3337-fig-02-a}
    \end{subfigure}
    \hfill	
    \begin{subfigure}{\figwidth}
        \includegraphics[width=\textwidth]{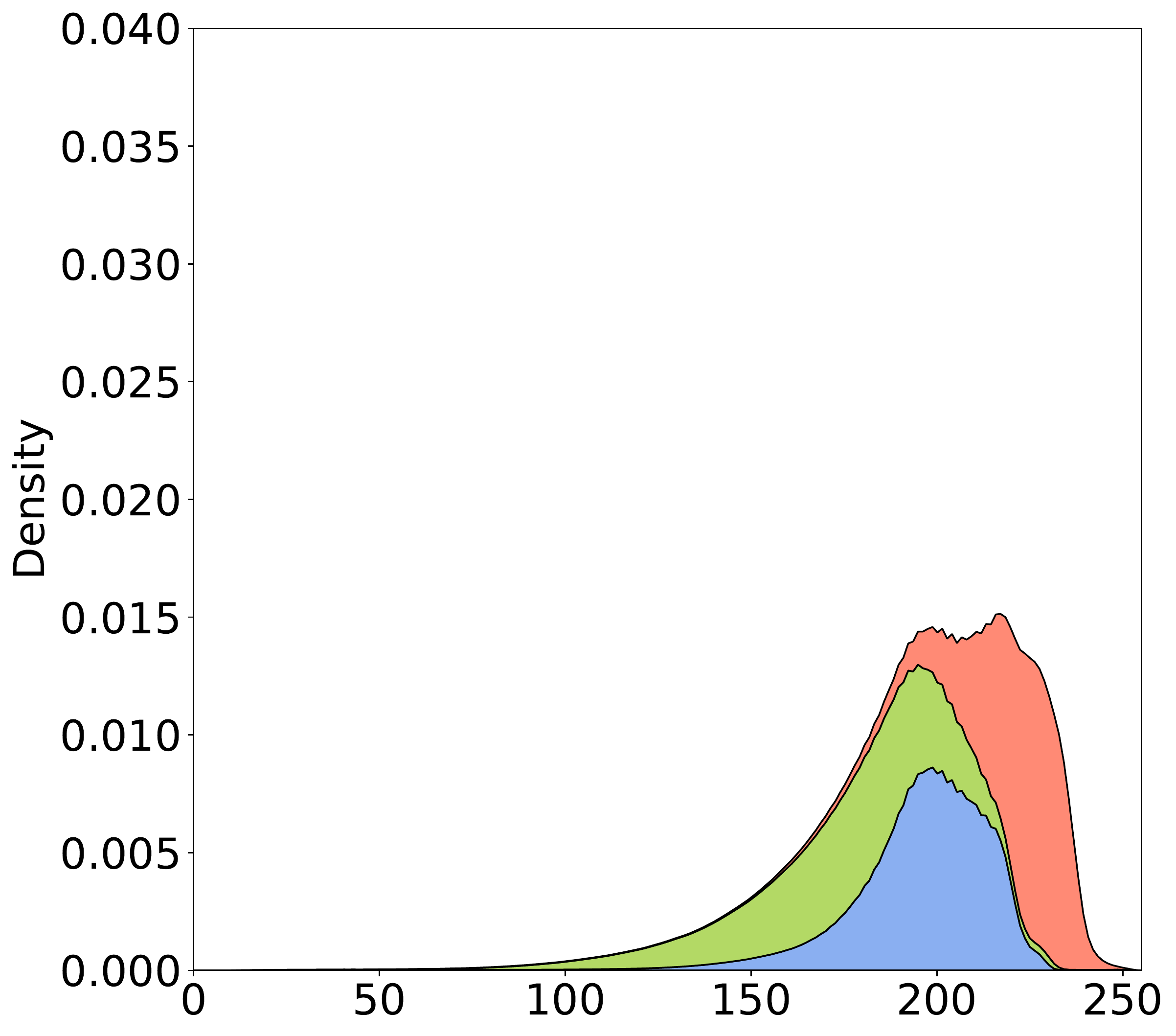}
        \caption{NZ210}
        \label{3337-fig-02-b}
    \end{subfigure}
    \hfill
    \begin{subfigure}{\figwidth}
        \includegraphics[width=\textwidth]{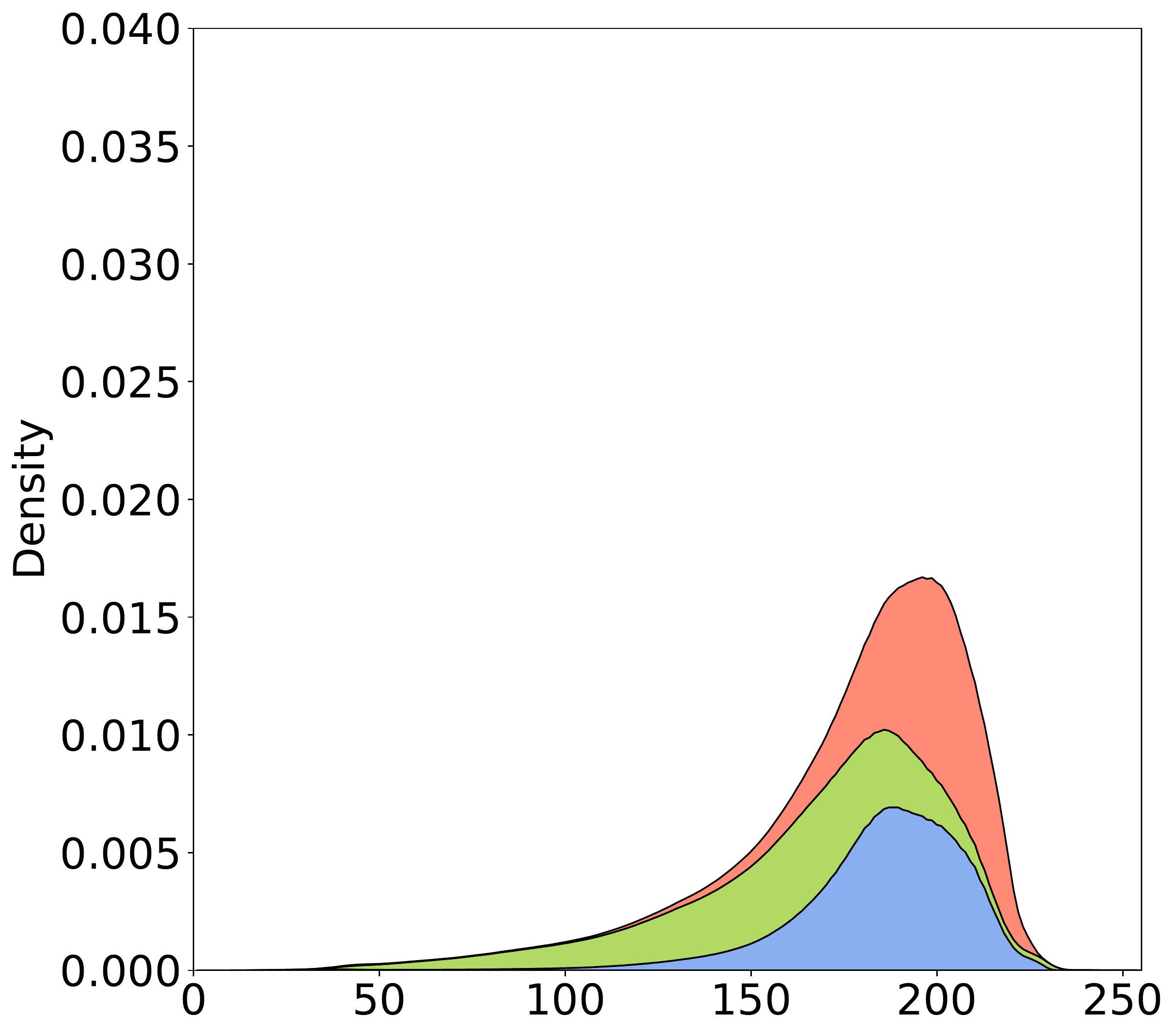}
        \caption{NZ2.0}
        \label{3337-fig-02-c}
    \end{subfigure} 
    \hfill
    \begin{subfigure}{\figwidth}
        \includegraphics[width=\textwidth]{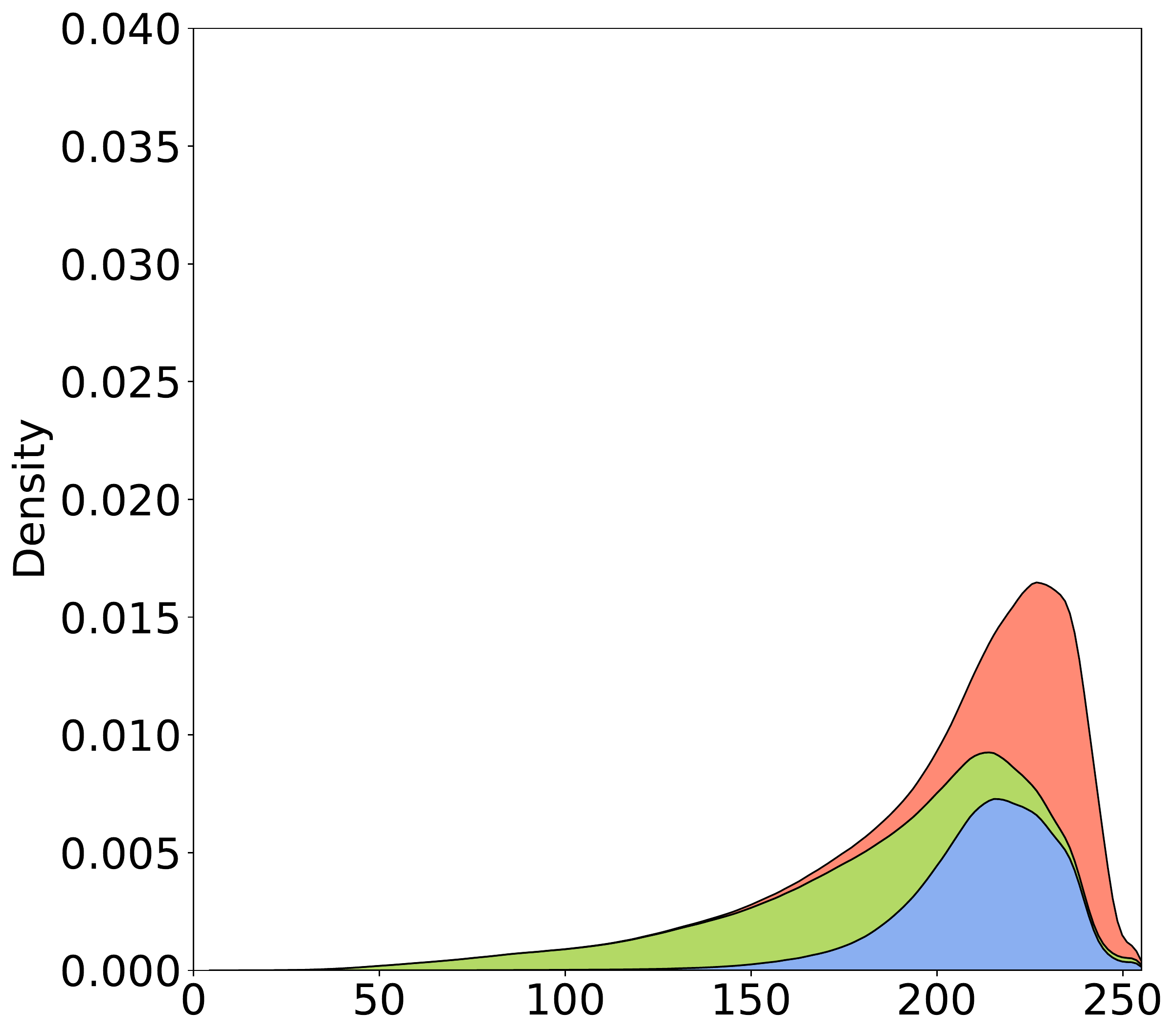}
        \caption{P1000}
        \label{3337-fig-02-d}
    \end{subfigure} 
    \hfill
    \begin{subfigure}{\figwidth}
        \includegraphics[width=\textwidth]{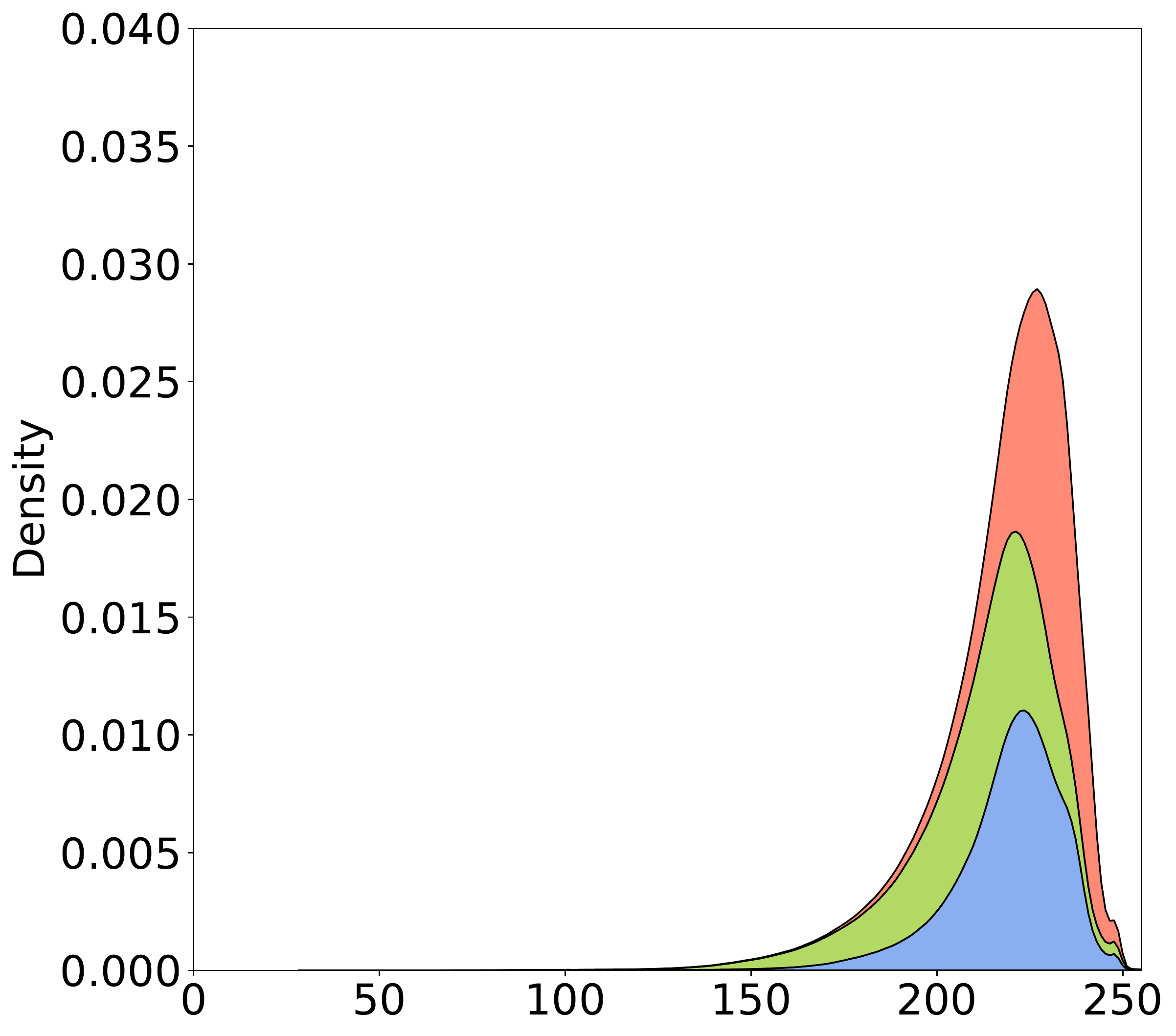}
        \caption{GT450}
        \label{3337-fig-02-e}
    \end{subfigure} 
\caption{Kernel density estimation of RGB values per scanner.}
\label{3337-fig-02}
\end{figure}

\begin{SCfigure}[10][h]
	\includegraphics[width=0.4\linewidth]{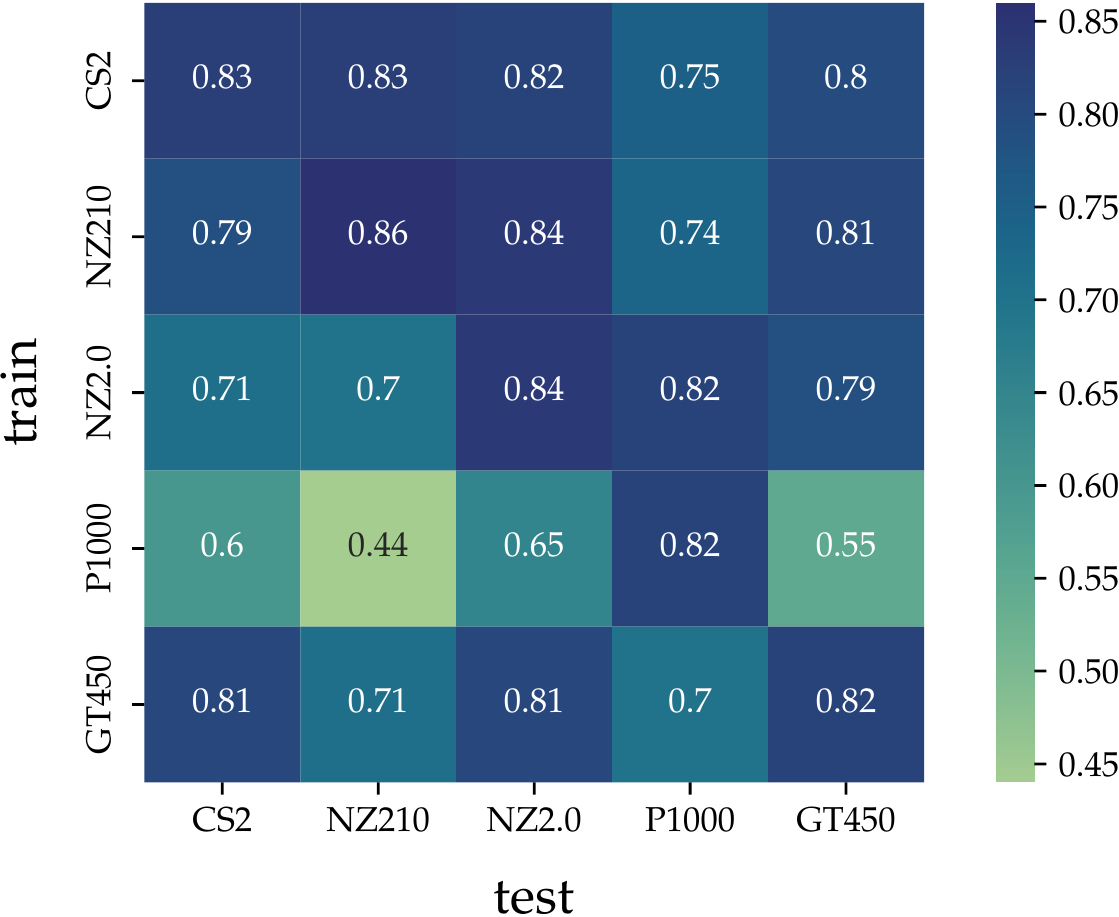}
	\caption{Scanner-wise performance of segmentation networks. Matrix entry $m_{i,j}$ is the mean intersection over union (mIoU) when training on the scanning system in row $i$ and testing on the scanning system in column $j$. Diagonal elements indicate in-domain performance, whereas off-diagonal elements represent cross-domain performance.}
	\label{3337-fig-03}
\end{SCfigure}

Figure~\ref{3337-fig-02} shows the RGB distribution of the tissue areas for the complete dataset of 44 WSIs per scanner. The distributions match the exemplary patches in Figure~\ref{3337-fig-01}, where the patches of the Aperio CS2 and the NanoZoomer 210 appear redder, reflected in a shift of the red pixel distributions to higher values. When looking at the distributions of the Aperio GT450, all curves are densely located at the higher color component values, which corresponds to the bright appearance of the patch in Figure~\ref{3337-fig-01-e}. Table~\ref{3337-tab-statistics} summarizes the channel-wise color averages, sharpness, and contrast of the slide scanning systems. These results further underline the visual impression of the patches in Figure~\ref{3337-fig-01}. When calculating the ratio of the red and the blue color channel $I_R/I_B$, the NZ210 results in a ratio of 1.12 and the NZ2.0 in a ratio of 1.04, which matches the much redder appearance of the NZ210 patch and the bluer appearance of the NZ2.0 patch. Overall, the CS2, NZ210, NZ2.0, and P1000 show comparable sharpness and contrast values, while the Aperio 450 exhibits a slightly higher sharpness but a considerably lower contrast.%
\begin{table}[t] 
\caption{Channel-wise color distributions $I_R, I_G,\text{ and }I_B$, sharpness $S_{CPBD}$ calculated as cumulative probability of blur detection, and Michelson contrast $C_M$ of the scanners ($\mu \pm \sigma$).}
\label{3337-tab-statistics}. 
\begin{tabular*}{\textwidth}{l@{\extracolsep\fill}ccccc}
    \hline
    & $I_R$ & $I_G$ & $I_B$ & $S_{CPBD}$ & $C_M$ \\ 
    \hline
    CS2 & 202.25 $\pm$ 10.22 & 153.83 $\pm$ 20.67 & 172.45 $\pm$ 16.64 & 0.80 $\pm$ 0.02 & 0.74 $\pm$ 0.12 \\
    NZ210 & 219.28 $\pm$ \phantom{0}8.12 & 173.27 $\pm$ 15.48 & 196.21 $\pm$ \phantom{0}9.82 & 0.82 $\pm$ 0.03 & 0.81 $\pm$ 0.14 \\
    NZ2.0 & 193.22 $\pm$ 10.77 & 154.34 $\pm$ 21.14 & 185.75 $\pm$ 11.58 & 0.81 $\pm$ 0.02 & 0.81 $\pm$ 0.13 \\
    P1000 & 223.56 $\pm$ 10.58 & 165.50 $\pm$ 24.80 & 212.15 $\pm$ 11.51 & 0.80 $\pm$ 0.02 & 0.71 $\pm$ 0.14 \\
    GT450 & 226.94 $\pm$ \phantom{0}6.65 & 208.67 $\pm$ 11.82 & 219.39 $\pm$ \phantom{0}8.03 & 0.84 $\pm$ 0.04 & 0.53 $\pm$ 0.15 \\
    \hline
\end{tabular*}	
\end{table}
Figure~\ref{3337-fig-03} visualizes the mIoU when training the segmentation network on one scanner, and testing it on all scanners. The results show high in-domain performance (diagonal) with mIoU values between 0.82 for the P1000 and GT450, and 0.86 for the NZ210. The cross-domain performance highlights the scanner-induced domain shift inherent in our dataset. While the networks trained on the CS2 and the NZ210 generalize considerably well, with performance decreases of only up to 0.08 and 0.12 compared to the in-domain mIoU, the highest cross-domain performance drop of up to 0.38 was observed when training on the P1000. The segmentation results showed that the network trained on the P1000 misclassified many background areas of the other scanners. A reason might be the integrated tissue detection of the P1000, which sets all pixels outside the tissue bounding box to (255, 255, 255) in order to reduce scanning times. This artificially removes common artifacts, e.g. dust particles, and the network might only look for high pixel values and not learn the morphological characteristics of background areas.        

\section{Discussion}
Our experiments have demonstrated the negative impact of scanner-induced domain shifts on the performance of neural networks, indicated by a considerable decrease in mIoU on unseen scanners. This confirms the observations of previous works and the need for methods that can tackle this domain shift and adequate datasets to evaluate their generalization capability. The presented dataset exceeds existing multi-scanner datasets regarding sample size and scanning systems. Furthermore, it provides local image correspondences, which isolate the scanner-induced from the morphology-induced domain shift and allow the development of algorithms dependent on these correspondences, e.g. WSI registration algorithms. We have implicitly shown the eligibility of our dataset for this application by successfully transferring the CS2 annotation database to the remaining scanners using WSI registration. The detailed evaluation of our scanner subsets has highlighted considerable differences in color distributions and contrasts present in clinically used scanners. Surprisingly, even though our evaluations resulted in the lowest contrast value for the Aperio GT450, this did not impede segmentation performance, shown by an in-domain mIoU of 0.82, which is comparable to the in-domain mIoUs of the remaining scanners. Our technical validation detected a large cross-domain performance decrease when training on the P1000 scanner. We assume that this can mainly be attributed to the unique pre-processing steps of the scanner vendor, as the P1000 showed similar image statistics to the CS2 but their average cross-domain performance differed considerably. However, we also observed a decrease in cross-domain performance for the remaining scanners, indicating that some of the learned feature representations did not generalize well across scanners. Future work could focus on a closer evaluation of which scanner characteristics hinder the extraction of domain-agnostic features and should therefore be disregarded, e.g. by using specific filters for data pre-processing or using adversarial training to punish the extraction of these features.    

\begin{acknowledgement}
F.W. gratefully acknowledges the financial support received by Merck Healthcare KGaA and the technical support received by the Clinical Assay Strategy 1 group at Merck Healthcare KGaA during sample digitization. K.B. gratefully acknowledges support by d.hip campus - Bavarian aim in form of a faculty endowment.
\end{acknowledgement}

\printbibliography

\end{document}